\documentstyle[prabib,aps,psfig,amsmath]{revtex}

\begin{document}

\title{An Approximate Model of the Spacetime Foam}
\author{V. Dzhunushaliev
\thanks{E-Mail Address : dzhun@rz.uni-potsdam.de and
dzhun@freenet.bishkek.su}}
\address{Universit\"at Potsdam, Institute f\"ur Mathematik,
14469, Potsdam, Germany \\
and Dept. Phys. and Microel. Engineer., Kyrgyz-Russian 
Slavic University\\ Bishkek, Kievskaya Str. 44, 720000, Kyrgyz
Republic}

\maketitle
\begin{abstract}
An approximate model of the spacetime foam is offered in which a
quantum handle (wormhole) is a 5D wormhole-like solution. Neglecting the 
linear sizes of the wormhole throat we can introduce a spinor
field for an approximate and effective description of the foam. 
The definition of the spinor field can be made by a dynamic and 
non-dynamic ways. In the first case some field equations are used 
and the second case leads to superspace. It
is shown that : the spacetime with the foam is similar
to a dielectric with dipoles and supergravity
theories with a non-minimal interaction between spinor and
electromagnetic fields can be considered as an effective model
for the spacetime foam.
\end{abstract}

\section{Introduction}

The notion of a spacetime foam was introduced by Wheeler \cite{wheel1} for
the description of the possible complex structure of the spacetime
on the Planck scale ($L_{Pl} \approx 10^{-33}cm$). The exact mathematical
description of this phenomenon is very difficult and even though
there is a doubt: does the Feynman path integral in the gravity contain
a topology change of the spacetime ? This question spring up as (according
to the Morse theory) the singular points must arise by topology changes. 
In such points the time arrow is undefined that leads
in difficulties at definition of the Lorentzian metric, curvature tensor
and so on.
\par
Here we propose an effective model of the spacetime foam
in which a spinor field is introduced for an approximate
description of the foam. For such model it is necessary the
nonminimal interaction between spinor and electromagnetic
fields (Pauli term). We will show that such interaction exists in the 5D
Kaluza-Klein theory with a spinor field in such a way
that the corresponding Maxwell equation is very similar
to the electrodynamic in the continuous media.
\par
\begin{figure}
\centerline{
\framebox{
\psfig{figure=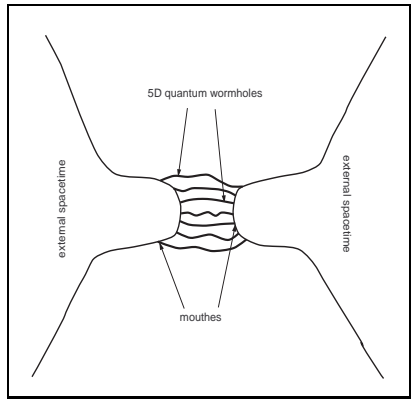,height=5cm,width=5cm}}}
\vspace{5mm}
\caption{The wormhole with a quantum throat. The cloud of QWHs can
be considered as the quantum throat.}
\label{fig1}
\end{figure}

In Ref.\cite{vd00b} is presented a model of the wormhole in which
a throat is a cloud of quantum wormholes (QWH)
(see Fig.(\ref{fig1})).
To describe these QWHs we introduce a spinor field.
In fact the spinor field is used for some
\textit{approximate and effective} description of QWHs as we
are not able to do it by a direct way. Here we would like to
offer the following model of the spacetime foam:
\begin{enumerate}
\item
Each QWH is a solution of the 5D vacuum Einstein equations with
$G_{5t}, G_{5\varphi} \neq 0$ components of the metric that 
leads to the appearance of electric and magnetic fields. In some 
approximation we can neglect of all linear sizes of QWH and obtain 
that Smolin \cite{smolin} calls as a ``minimalist'' wormhole. 
For the 4D observer each mouth look as
$(\pm)$ electric charge since this mouth entraps the force lines
of the electric field (see Fig.(\ref{fig3})).
\item
For the external observer each QWH is like to dipole and the
spacetime with the foam seems as a dielectric by filled dipoles
(see Fig.(\ref{fig2})).
\item
The spacetime foam is described by a spinor field $\psi$ and
the physical meaning of $\psi$ depends on an interaction term
between spinor and electromagnetic fields that we shall discuss
below.
\item
An interaction between electromagnetic and spinor fields
is nonminimal that allows us to interpret the Maxwell
equations like to the electrodynamic in a continuous media.
\begin{figure}
\centerline{
\framebox{
\psfig{figure=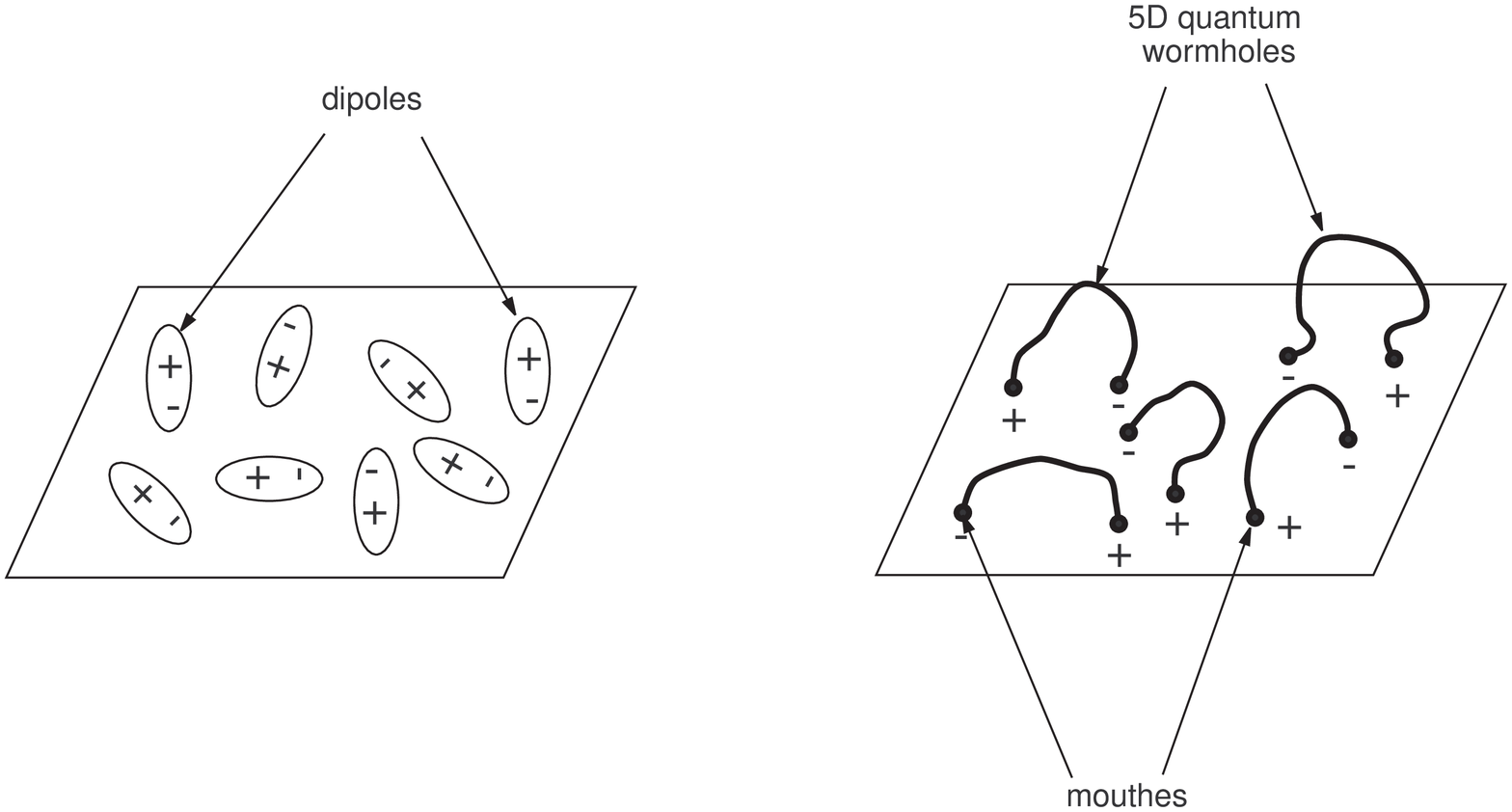,height=5cm,width=10cm}}}
\vspace{5mm}
\caption{For the 4D observer each mouth looks as a moving electric
charge. This allows us in some approximation imagine the
spacetime foam as a continuous media with a polarization.}
\label{fig2}
\end{figure}
\end{enumerate}

\section{Model of the individual quantum wormhole}

\begin{figure}
\centerline{
\framebox{
\psfig{figure=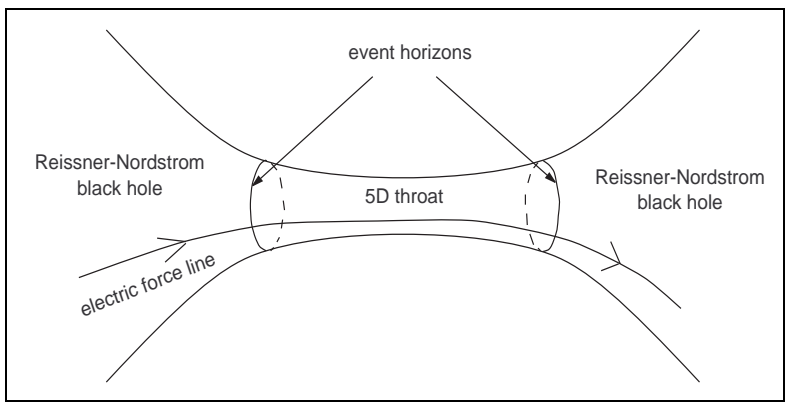,height=5cm,width=10cm}}}
\vspace{5mm}
\caption{The model of the individual quantum wormhole.
The whole spacetime is 5 dimensional but: in the
Reissner-Nordstr\"om black hole $G_{55} = const$ and it is not
varying (this is the 5D gravity in the initial Kaluza-Klein
interpretation); in the 5D throat $G_{55}$ is the dynamical
variable and we have 15 equations which are equivalent to
4D Einstein + Maxwell + scalar equations.}
\label{fig3}
\end{figure}

The model of the individual QWH is presented on the Fig.(\ref{fig3}).
In fact this is some realization of the Wheeler idea about
a wormhole entrapping electric force lines. In Ref.\cite{wheel1}
he wrote: ``Along with the fluctuations in the metric there occur
fluctuations in the electromagnetic field. In consequence
the typical multiply connected space $\ldots$ has a net flux of
electric lines of force passing through the "wormhole". These
lines are trapped by the topology of the space. These lines give
the appearance of a positive charge at one end of the wormhole
and a negative charge at the other''.
The composite wormhole on the Fig.(\ref{fig3}) consists
from two Reissner-Nordstr\"om black holes and the 5D throat
inserted between them \cite{dzh7}. The 5D metric for this throat
is
\begin{equation}
ds^{2}_{(5)} = - R_0^2 e^{2\psi(r)}\Delta (r)
\left(d\chi  + \omega (r)dt + Q \cos \theta d\varphi \right)^2 + 
\frac{1}{\Delta (r)}dt^{2} - dr^{2} - a(r)
\left (
d\theta ^2 + \sin\theta ^2 d\varphi ^2
\right ),
\label{ind1}
\end{equation}
where $\chi $ is the 5$^{th}$ extra coordinate;
$R_0 > 0$ and $Q$ are some constants. And we assume that 
in some approximation such QWH of spacetime foam can be presented 
by this manner. 
\par
The 5D Einstein equations are 
\begin{eqnarray}
  \frac{\Delta ''}{\Delta} - \frac{{\Delta'}^2}{\Delta^2} + 
  \frac{\Delta' \psi'}{\Delta} + \frac{a' \Delta'}{a \Delta} + 
  R_0^2{\omega'}^2 \Delta^2 e^{2\psi} & = & 0 ,
\label{ind1a}\\
  \omega '' + \omega'\left( 2\frac{\Delta'}{\Delta} + 3\psi' + 
  \frac{a'}{a}\right) & = & 0 ,
\label{ind1b}\\
  \frac{a''}{a}   + \frac{a'\psi'}{a} - \frac{2}{a} + 
  \frac{Q^2\Delta e^{2\psi}}{a^2} & = & 0 ,
\label{ind1c}  \\
  \psi'' + {\psi'}^2 + \frac{a'\psi'}{a} - 
  \frac{Q^2\Delta e^{2\psi}}{2a^2} & = & 0 ,
\label{ind1d}  \\
  \frac{{\Delta'}^2}{\Delta^2} + 2\frac{\Delta'\psi'}{\Delta} - 
  4\frac{a'\psi'}{a} + \frac{4}{a} - \frac{{a'}^2}{a^2} - 
  R_0^2{\omega'}^2 \Delta^2 e^{2\psi} - 
  \frac{Q^2\Delta e^{2\psi}}{a^2}& = & 0 .
\label{ind1e}  
\end{eqnarray}
In Ref. \cite{dzhsin} it is shown that there is three type of solutions : 
the first type (wormhole-like solution) is presented on Fig.(\ref{fig3}) 
with $E > H$ ($E$ and $H$ are Kaluza-Klein electric and magnetic fields), 
the second one is an infinite flux tube with $E = H$ and the third one 
is a singular solution (finite flux tube) with $E < H$. The definitions 
for $E$ and $H$ fields will be given later. 
\par 
The longitudinal size $l_0$ of the WH-like solution depends on the 
relation between $E$ and $H$ : if $H/E \rightarrow 1$ then 
$l_0 \rightarrow \infty$. Let us define an approximate solution 
close to points $r^2 = r^2_0$ (where $ds^2(\pm r_0) = 0$). This solution 
we search in the form 
\begin{eqnarray}
  \Delta & \approx & \Delta_1 \left( r_0^2 - r^2 \right) , 
\label{ind1f}\\
  \omega & \approx & \frac{\omega_1}{r_0^2 - r^2} ,
\label{ind1g}\\
  \psi & \approx & \frac{\psi_3}{6} \left( r_0^2 - r^2 \right)^3 .
\label{ind1h}  
\end{eqnarray}
The solution is 
\begin{eqnarray}
  \Delta_1 & = & \pm \frac{q}{2a_0r_0} , 
\label{ind1j}\\
  \omega_1 & = & \frac{2 a_0 r_0}{q} ,
\label{ind1k}\\
  \psi_3 & = & \pm \frac{qQ^2}{2a_0^3 r_0^3} 
\label{ind1l}  
\end{eqnarray}
here $a_0 = a(r=\pm r_0)$, $q$ is some constant. It is easy to show 
that at the hypersurfaces $r = \pm r_0$ : $ds^2 = 0$. On these 
hypersurfaces the change of the metric signature takes place : 
$(+,-,-,-,-)$ by $|r| < r_0$ and $(-,-,-,-,+)$ by 
$|r| > r_0$. Following to Bronnikov \cite{bron} we call these 
two hypersurfaces as $T-$horizons. 
\par 
For the definition of a Kaluza-Klein electric field we 
consider Eq.(\ref{ind1b}) 
\begin{equation}
  \left[\left(\omega' \Delta^2 e^{3\psi}\right) 
  4\pi a  
  \right]' = 0 
\label{ind1q}
\end{equation}
here $4\pi a$ is the area of $S^2$ sphere. 
Comparing with the Gauss law we see that Kaluza-Klein electric 
field can be defined as follows 
\begin{equation}
  E_{KK} = \omega' \Delta^2 e^{3\psi} = \frac{q}{a}
\label{ind1w}
\end{equation}
here $q$ is an electric charge which is proportional to a flux of 
electric field. In this case the force lines of the electric field 
are uninterrupted and can be continued through the surfaces of matching 
the 5D WH-like solution and the Reissner-Nordstr\"om solution like to 
Fig.\ref{fig3}. For the definition of a Kaluza-Klein magnetic field 
we write the following 5D Einstein equation 
\begin{equation}
  R_{\chi\varphi} = \frac{e^\psi \sqrt{\Delta}}{a^2 \sin \theta} 
  \frac{\partial}{\partial\theta} \biggl( Q \biggl) = 0 
\label{ind1r}
\end{equation}
and compare it with the ordinary 4D Maxwell equation 
\begin{equation}
  \frac{1}{\sqrt{-g}} \frac{\partial}{\partial x^\mu}
  \biggl(\sqrt{-g} F^{\varphi\mu}\biggl) = \frac{1}{\sin\theta}
  \frac{\partial}{\partial \theta} 
  \biggl(\sin\theta F^{\theta\varphi} \biggl) = 
  \frac{1}{\sin\theta}
  \frac{\partial}{\partial \theta} 
  \biggl( \frac{H_r}{a} \biggl) = 0 
\label{inf1t}
\end{equation}
here $F^{\theta\varphi} = 
\frac{\epsilon^{\theta\varphi i}}{\sqrt{\gamma}}H_i$, 
$\varepsilon^{ijk}$ is the antisymmetrical tensor, 
$\gamma$ is the determinant of the 3D metric. The result is 
\begin{equation}
  H_r = \frac{Q\sqrt{\Delta} e^\psi}{a}.
\label{ind1y}
\end{equation}
Immediately we see that $H_r \rightarrow 0$ by $r \rightarrow \pm r_0$. 
The Einstein equations tell us that close to hypersurface 
$r = \pm r_0$ the Kaluza-Klein 
magnetic field can not have any influence on the gravity 
as the following term in Eq's \eqref{ind1c}, \eqref{ind1d} and 
\eqref{ind1e} tends to zero 
\begin{equation}
    H_r^2 = \frac{Q^2 \Delta e^{2\psi}}{a^2} \rightarrow 0
    \quad \text{by} \quad r \rightarrow \pm r_0. 
\label{ind1u}
\end{equation}
It means that the WH-like solutions near to these hypersurfaces 
are identical to the solution without the magnetic field. The external 
4D observer sees that the force lines of magnetic field do not 
cross the event horizon for such composite WH. Another words 
each of QWH is like to moving electric charge but not a magnetic charge. 
\par
On these $T-$horizons we should match:
\begin{itemize}
\item
the flux of the 4D electric field (defined by the Maxwell
equations) with the flux of the 5D electric field defined
by $R_{5t} = 0$ Kaluza-Klein equation.
\item
the area of the Reissner-Nordstr\"om event horizon with
the area of the $T-$hprizon.
\end{itemize}
It is necessary to note that both solutions (Reissner-Nordstr\"om
black hole and 5D throat) have only two integration
constants\footnote{in fact, for the Reissner-Nordstr\"om
black hole this leads to the ``no hair'' theorem.} and
on the event horizon takes place an algebraic relation
between these 4D and 5D integration constants. Another
explanation of the fact that we use only two joining condition
is the following (see Ref.\cite{dzh99yb} for the more detailed
explanations): in some sense on the event horizon holds
a ``holography principle''. This means that in the presence
of the event horizon the 4D and 5D Einstein equations lead
to a reduction of the amount of initial data. For example
the Einstein - Maxwell equations for the Reissner-Nordstr\"om
metric
\begin{eqnarray}
ds^2 & = & \Delta dt^2 - \frac{dr^2}{\Delta} - r^2
\left (
d\theta ^2 + \sin^2 d\varphi ^2
 \right ) ,
\label{ind21}\\
A_\mu & = & \left (
\omega ,0,0,0)
 \right )
\label{ind22}
\end{eqnarray}
(where $A_\mu$ is the electromagnetic potential, $\kappa$
is the gravitational constant) can be written as
\begin{eqnarray}
-\frac{\Delta '}{r} + \frac{1 - \Delta}{r^2} & = &
\frac{\kappa}{2} {\omega '} ^2  ,
\label{ind3} \\
\omega ' & = & \frac{q}{r^2}     .
\label{ind4}
\end{eqnarray}
For the Reissner - Nordstr\"om black hole the event horizon
is defined by the condition $\Delta (r_g) = 0$, where $r_g$
is the radius of the event horizon.
Hence in this case we see that on the event horizon
\begin{equation}
\Delta '_g = \frac{1}{r_g} - \frac{\kappa}{2}
r_g {\omega '_g}^2  ,
\label{ind5}
\end{equation}
here (g) means that the corresponding value is taken
on the event horizon. Thus, Eq. (\ref{ind3}), which is the
Einstein equation, is the first-order differential equation 
in the whole spacetime $(r \ge r_g)$. The condition (\ref{ind5})
tells us that the derivative of the metric on the event horizon
is expressed through the metric value on the event horizon.
This is the same what we said above: the reduction of the
amount of initial data takes place by such a way that we have
only two integration constants (mass $m$ and charge $e$ for the
Reissner-Nordstr\"om solution and $q$ and $r_0$ for the 5D
throat).

\section{Spacetime Foam and Spinor Fields}
\label{secspinor}

\begin{figure}
\begin{center}
\fbox{
\psfig{figure=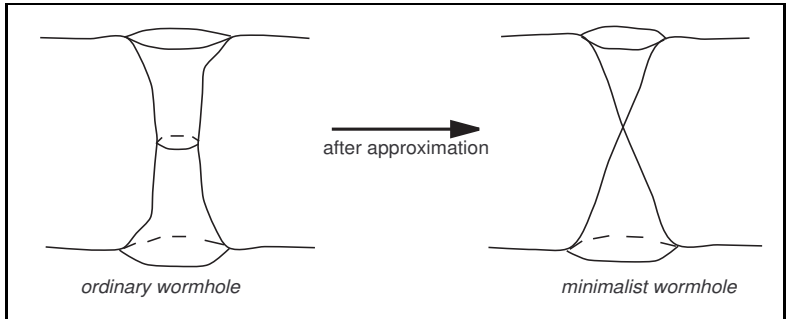,height=4cm,width=10cm}}
\caption{The minimalist wormhole.}
\label{fig1a}
\end{center}
\end{figure}

On the next approximation step we want to neglect with a cross section 
and longitudinal length of the 5D throat. In the result each QWH 
looks as an identification of two points, see Fig.\ref{fig1a}
Following to Smolin \cite{smolin} we introduce 
an operator $\hat{A}^{ab}(x^\mu,y^\mu)$ describing a quantum 
state in which the space with two points $x$ and $y$ fluctuates 
between two possibilities : points $(x,y)$ \textbf{either} are 
pasted together \textbf{or} not. In fact this operator describes 
an undeterminacy connected with the creation/annihilation of a 
wormhole\footnote{This is like to spin : $z-$projection 
of spin can have two values $\pm\hbar/2$.}. 
Smolin calls such wormhole as a minimalist wormhole. 
The minimalist wormhole can be received from the 
above-mentioned composite wormhole if we neglect the linear sizes 
of 5D throat, \textit{i.e.} shrink their to a point 
(see, Fig.\ref{fig1a}). 
We demand that this operator 
$\hat{A}^{ab}(x,y)$ should have the following property 
\begin{equation}
    \hat{A}^{ab}(x,y) = \theta^a(x) \theta^b(y)
\label{sec3.2-2}
\end{equation}
$a,b$ are some indices which will be determine later. 
Of coarse for the definition of $\theta^a(x)$ we should have some 
additional equation for this quantity. Smolin's definition is 
\begin{eqnarray}
  \hat{A}^{ab}(x,y) & = & \epsilon^{ab} \hat{A}(x,y) = 
  \theta^a(x) \theta^b(y) ,
\label{sec3.2-3}\\
  \left(\hat{A}(x,y)\right)^2 & = & 0 
\label{sec3.2-4}
\end{eqnarray}
here $a,b$ are the spinor indices and $\theta^a(x)$ is a spinor, 
$\epsilon^{12} = -\epsilon^{21} = 1, 
\epsilon^{11} = \epsilon^{22} = 0$. 
We would like to say that it can be various definitions : 
a dynamical definition with  
field equations is given in section \ref{effective} and the definition 
for which $\theta^a(x) = \theta^a = const$ will be given in section 
\ref{supergravity}.

\section{An effective model of the spacetime foam}
\label{effective}

In this case the quantity $\theta^a(x)$ is a spinor field 
$\psi^\alpha(x) = \theta^\alpha(x)$ ($a = \alpha$ is the spinor index) 
and we determine $\psi(x)$ dynamically by means of some 
field equations for $\psi(x)$.
\par
It is well known that gauge fields naturally appears in
multidimensional gravities \cite{sal1}, \cite{per1}, \cite{coq1983zn}
as some components of the multi-bein.
But for the
spinor field it is not the case. The spinor field in 4D and 5D
spacetimes can have different interaction with gauge fields.
For the 4D case the interaction term in Lagrangian
is minimal $(i\bar\psi A_\mu \gamma^\mu \psi$, 
$\mu$ is the 4D index) but for the
second case it can be
$(i\bar\psi F_{AB}\gamma^A \gamma^B \psi)$ or
$(iF_{AB} \bar\psi^A \psi^B)$ (here $\psi^A$ is
the Rarita-Schwinger spinor, $A,B$ are the 5D indices) 
or something like this.
\par
We will consider the 5D Kaluza-Klein theory + torsion +
spinor field with the 5D metric  
\begin{equation}
  ds^2 = - \left( d\chi + A_\mu dx^\mu \right)^2 + 
  g_{\mu\nu} dx^\mu dx^\nu 
\label{md0}
\end{equation}
In this case (according to the initial interpretation of the 
Kaluza-Klein gravity with 
$G_{55} = const$) we have the electromagnetic potential $A_\mu$ and 
the 4D metric $g_{\mu\nu}$. The Lagrangian for this theory is
\begin{eqnarray}
{\cal L} & = & \sqrt{-G}
\left \{
-\frac{1}{2k}
  \left (
  R^{(5)} - S_{ABC}S^{ABC}
  \right ) +
\right.
\nonumber \\
&&
\left.
 \frac{\hbar c}{2}
  \left \{
  i\bar\psi \left [\gamma^C
        \left (
        \partial_C - \frac{1}{4}\omega _{\bar A \bar B C}
        \gamma^{[\bar A}\gamma^{\bar B]} -
        \frac{1}{4}S_{\bar A \bar B C}
        \gamma^{[\bar A}\gamma^{\bar B]} \right ) - \frac{mc}{i\hbar}
        \right ]\psi  + h.c.
  \right \}
\right\}
\label{md1}
\end{eqnarray}
where $G$ is the determinant of the 5D metric,
$R^{(5)}$ is the 5D scalar curvature, $S_{ABC}$ is the
antisymmetrical torsion tensor, $A,B,C$ are the 5D world indexes,
$\bar A, \bar B, \bar C$ are the 5-bein indexes,
$\gamma ^B = h^B_{\bar A}\gamma^{\bar A}$,
$h^B_{\bar A}$ is the 5-bein,
$\gamma^{\bar A}$ are the 5D $\gamma$ matrixes with usual definitions
$\gamma^{\bar A}\gamma^{\bar B} + \gamma^{\bar B}\gamma^{\bar A} =
2\eta^{\bar A\bar B}$,
$\eta^{\bar A\bar B} = (+,-,-,-,-)$ is the signature of the
5D metric; $[]$ means the antisymmetrization, 
$\hbar$, $c$ and $m$ are the usual constants.
The most important for us is the choice of a spinor $\psi$ 
which will approximately describe the spacetime foam as it was 
mentioned in the section \ref{secspinor} : \textit{i.e.} 
$\psi^\alpha (x)= \theta^\alpha (x)$. It is very important to note 
that in the context of this section we have some dynamical 
equations for the 
$\theta^\alpha(x) = \psi^\alpha(x)$ (it is convenient to use 
the usual designation $\psi$ for the fermion field). 
We should note that all 
physical fields in Lagrangian \eqref{md1} must be quantum operators 
but on the first approximation step we change their by 
classical fields. After dimensional reduction
we have
\begin{eqnarray}
{\cal L} & = & \sqrt{-g}
\left \{
-\frac{1}{2k}
        \left (
        R + \frac{1}{4}F_{\alpha\beta}F^{\alpha\beta}
        \right ) +
\right.
\nonumber \\
&&
\left.
\frac{\hbar c}{2}
        \left [
        i\bar \psi
                \left (
                \gamma^\mu \tilde\nabla_\mu -
                \frac{1}{8}F_{\bar\alpha\bar\beta}\gamma^{\bar 5}
                \gamma^{[\bar\alpha}\gamma^{\bar\beta ]} -
                \frac{1}{4}l^2_{Pl}
                \left (\gamma^{[\bar A}\gamma^{\bar B}\gamma^{\bar C]} \right )
                \left (i\bar\psi \gamma_{[\bar A}\gamma_{\bar B}\gamma_{\bar C]}
                \psi \right ) - \frac{mc}{i\hbar}
                \right )\psi  + h.c.
        \right ]
\right \}
\label{md2}\\
S^{\bar A\bar B\bar C} & = &
2l^2_{Pl}
\left (
i\bar\psi \gamma^{[\bar A}\gamma^{\bar B}\gamma^{\bar C]}\psi
\right )
\label{md3}
\end{eqnarray}
where $g$ is the determinant of the 4D metric,
$R$ is the 4D scalar curvature,
$F_{\alpha\beta} = \partial_\alpha A_\beta - \partial_\beta A_\alpha$
is the Maxwell tensor,
$A_\mu = h^{\bar 5}_\mu$ is the electromagnetic potential,
$\alpha ,\beta ,\mu$ are the 4D world indexes,
$\bar\alpha ,\bar\beta, \bar\mu$ are the 4D vier-bein indexes,
$h^{\bar \mu}_\nu$ is the vier-bein,
$\gamma^{\bar \mu}$ are the 4D $\gamma$ matrixes with usual definitions
$\gamma^{\bar\mu}\gamma^{\bar\nu} + \gamma^{\bar\nu}\gamma^{\bar\mu} =
2\eta^{\bar\mu\bar\nu}$,
$\eta^{\bar\mu\bar\nu} = (+,-,-,-)$ is the signature of the
4D metric. Varying with respect to $g_{\mu\nu}$, $\bar\psi$
and $A_\mu$ leads to the following equations
\begin{eqnarray}
R_{\mu\nu} - \frac{1}{2}g_{\mu\nu}R =
\frac{1}{2}
\left (
-F_{\mu\alpha}F_\nu^\alpha + \frac{1}{4}g_{\mu\nu}
F_{\alpha\beta} F^{\alpha\beta}
 \right ) & + &
4l^2_{Pl}
\left [
        \left (
        i\bar\psi\gamma_\mu \tilde\nabla_\nu \psi +
        i\bar\psi\gamma_\nu \tilde\nabla_\mu \psi
         \right ) + h.c.
 \right ] -
\nonumber \\
2l^2_{Pl}
\left [
F_{\mu\alpha}
\left (i\bar\psi \gamma^{\bar 5} \gamma_{[\nu}
\gamma^{\alpha ]} \psi \right ) +
F_{\nu\alpha}
\left (i\bar\psi \gamma^{\bar 5} \gamma_{[\mu}
\gamma^{\alpha ]} \psi \right )
 \right ] & - & 
2g_{\mu\nu} l^4_{Pl}
\left (
i\bar\psi \gamma^{[\bar A}\gamma^{\bar B}\gamma^{\bar C]}\psi
\right )
\left (
i\bar\psi \gamma_{[\bar A}\gamma_{\bar B}\gamma_{\bar C]}\psi
\right ) ,
\label{md4}\\
D_\nu H^{\mu\nu} = 0 ,
\,
H^{\mu\nu} =
F^{\mu\nu} + \tilde F^{\mu\nu} ;
\,
\tilde F^{\mu\nu} & = & 4l^2_{Pl}
\left (
i\bar\psi \gamma^{\bar 5}\gamma^{[\mu} \gamma^{\nu ]}\psi
 \right ) =
4l^2_{Pl}
E^{\mu\nu\alpha\beta}
\left (
i\bar\psi \gamma_{[\alpha} \gamma_{\beta ]}\psi
 \right ) ,
\label{md5}\\
  i\gamma^\mu \tilde \nabla _\mu \psi-
  \frac{1}{8} F_{\bar \alpha \bar \beta}
  \left (
     i\gamma^{\bar 5}\gamma^{[\bar \alpha} \gamma^{\bar \beta ]} \psi
  \right ) & - &
  \frac{1}{2} l^2_{Pl}
  \left (
     i\gamma^{[\bar A}\gamma^{\bar B}\gamma^{\bar C]} \psi
  \right )
  \left (
     i\bar\psi \gamma_{[\bar A}\gamma_{\bar B}\gamma_{\bar C]}\psi
  \right ) = 0 ,
\label{md6}\\
\tilde \nabla _\mu & = &
\partial _\mu - \frac{1}{4} \omega _{\bar a\bar b \mu}
\gamma^{[\bar a} \gamma^{\bar b]}
\label{md7}
\end{eqnarray}
where $\tilde \nabla _\mu$ is the 4D covariant derivative
of the spinor field without torsion,
$\omega _{\bar a\bar b \mu}$ is the 4D Ricci coefficients without
torsion, $E^{\mu\nu\alpha\beta}$ is the 4D absolutely antisymmetric
tensor. The most interesting for us is the Maxwell equation
(\ref{md5}) which permits us to discuss the physical meaning
of the spinor field.
We would like to show that this equation in the given form
is similar to the electrodynamic in the continuous media.
Let we remind that for the electrodynamic in the continuous media
two tensors $\bar F^{\mu\nu}$ and $\bar H^{\mu\nu}$ are introduced
\cite{landau} for which we have the following equations system
(in the Minkowski spacetime)
\begin{eqnarray}
\bar F_{\alpha\beta ,\gamma} +
\bar F_{\gamma\alpha ,\beta} +
\bar F_{\beta\gamma ,\alpha} & = & 0 ,
\label{md8}\\
\bar H^{\alpha\beta}_{,\beta} & = & 0
\label{md9}
\end{eqnarray}
and the following relations between these tensors
\begin{eqnarray}
\bar H_{\alpha\beta} u^\beta & = &
\varepsilon \bar F_{\alpha\beta}u^\beta ,
\label{nd10}\\
\bar F_{\alpha\beta} u_\gamma +
\bar F_{\gamma\alpha} u_\beta +
\bar F_{\beta\gamma} u_\alpha & = &
\mu
\left (
\bar H_{\alpha\beta} u_\gamma +
\bar H_{\gamma\alpha} u_\beta +
\bar H_{\beta\gamma} u_\alpha
 \right )
\label{md11}
\end{eqnarray}
where $\varepsilon$ and $\mu$ are the dielectric
and magnetic permeability respectively, $u^\alpha$ is the
4-vector of the matter. For the
rest media and in the 3D designation we have
\begin{eqnarray}
\varepsilon \bar E_i & = & \bar E_i + 4\pi \bar P_i = \bar D_i ,
\quad \mbox{where} \quad
\bar E_i = \bar F_{0i} ,
\quad
\bar D_{0i} = \bar H_{0i} ,
\label{md12}\\
\mu \bar H_i & = &
\bar H_i + 4\pi \bar M_i = \bar B_i ,
\quad \mbox{where} \quad
\bar B_i = \epsilon_{ijk}\bar F^{jk} ,
\quad
\bar H_i = \epsilon_{ijk}\bar H^{jk} ,
\label{md13}
\end{eqnarray}
where $P_i$ is the dielectric polarization and $M_i$ is the
magnetization vectors, $\epsilon_{ijk}$ is the 3D absolutely
antisymmetric tensor. Let us rewrite the definition in 
Eq.\eqref{md5} in the following form 
\begin{eqnarray}
E_i + \tilde E_i & = & D_i
\quad \mbox{where }\quad
E_i = F_{0i}, \quad \tilde E_i = \tilde F_{0i} ,
\quad D_i = H_{0i}
\label{md14}\\
B_i + \tilde B_i & = & H_i
\quad \mbox{where} \quad
B_i = \epsilon_{ijk} F^{jk},
\quad
\tilde B_i = \epsilon_{ijk} \tilde F^{jk},
\quad
H_i = \epsilon_{ijk} H^{jk}.
\label{md15}
\end{eqnarray}
Comparing these definitions with \eqref{md12}, \eqref{md13} 
immediately 
we see that the following notations can be introduced 
\begin{equation}
\tilde E_i = 4l^2_{Pl} \epsilon_{ijk}
\left (
i\bar \psi \gamma^{[j} \gamma^{k]} \psi
 \right )
\label{md16}
\end{equation}
is the polarization vector of the spacetime foam and
\begin{equation}
\tilde B_i = -4l^2_{Pl} \epsilon_{ijk}
\left (
i\bar \psi \gamma ^{\bar 5} \gamma^{[j} \gamma^{k]} \psi
 \right )
\label{md17}
\end{equation}
is the magnetization vector of the spacetime foam.
\par
The physical reason for this is : each QWH is like
to a moving dipole (see Fig.(\ref{fig2})) which produces
microscopical electric and magnetic fields.

\section{Supergravity as a possible model of the spacetime foam}
\label{supergravity}

Now we would like to consider the another possible 
interpretation of $\theta^a (x)$. 
In this case (in contrast to section \ref{effective}) 
we determine the quantity $\theta^a(x)$ by a non-dynamic 
way : introducing some algebraic equation \eqref{sec5.2-1} for 
$\theta$. Let the operator $\hat{A}^{ab}(x,y)$ 
introduced in the section \ref{secspinor} satisfies to the 
following equation 
\begin{equation}
    \hat{A}^{ab}(x,y) \hat{A}_{ab}(x,y) = 0 .
    \label{sec5.2-1}
\end{equation}
The solution of Eq.\eqref{sec5.2-1} we search in the form 
\begin{equation}
    \hat{A}^{ab}(x,y) = \theta^a(x) \theta^b(y).
\label{sec5.2-2}
\end{equation}
After substitution in Eq.\eqref{sec5.2-1} we have 
\begin{equation}
  \theta^a(x) \theta^b(y) \theta_a(x) \theta_b(y) = 0.
\label{sec5.2-3}
\end{equation}
Like to section \ref{secspinor} we have two simplest solution. 
The first solution is 
\begin{eqnarray}
    a & = & \alpha, \; b = \beta , \; (\alpha , \beta= 1,2) ,
    \label{sec5.2-0a}\\
    \hat{A}^{\alpha\beta}(x,y) & = & 
    \theta^\alpha (x) \theta^\beta (y).
    \label{sec5.2-0c}
\end{eqnarray}
$\theta^\alpha$ is an undotted spinor of 
$(\frac{1}{2},0)$ representation of Sl(2,\textbf{C}) group. 
\begin{eqnarray}
    \theta^\alpha(x) & = & \theta^\alpha(y) = \theta^\alpha = const ,
    \label{sec5.2-4}\\
    \theta^\alpha \theta^\beta & = & - \theta^\beta \theta^\alpha ; 
    \qquad \alpha \neq \beta ,
    \label{sec5.2-5}\\
    \left(\theta^\alpha \right)^2 & = & 0 .
    \label{sec5.2-6}
   \end{eqnarray}
The second solution is similar : 
$\alpha \rightarrow \dot \alpha$ and 
$\beta \rightarrow \dot \beta$ 
\begin{eqnarray}
    a & = & \dot\alpha, \; b = \dot\beta , \; 
    (\dot\alpha , \dot\beta= 1,2) , 
\label{ces5.2-7}\\
    \hat{A}^{\dot\alpha\dot\beta}(x,y) & = & 
    \bar\theta^{\dot\alpha} (x) \bar{\theta}^{\dot\beta} (y),
\label{sec5.2-7a}\\
    \bar\theta^{\dot\alpha}(x) & = & \bar\theta^{\dot\alpha}(y) = 
    \bar\theta^{\dot\alpha} = const ,
\label{sec5.2-8}\\
    \bar\theta^{\dot\alpha} \bar\theta^{\dot\beta} & = & 
    - \bar\theta^{\dot\beta} \bar\theta^{\dot\alpha} ; 
    \qquad \dot\alpha \neq \dot\beta ,
\label{sec5.2-9}\\
    \left(\bar\theta^{\dot\alpha} \right)^2 & = & 0 .
\label{sec5.2-10}
\end{eqnarray}
In this case $\bar\theta^{\dot\alpha}$ is a dotted spinor of 
$(0,\frac{1}{2})$ representation. 
Such two-valuedness compels us to introduce both possibilities : 
$\theta = \{ \theta^\alpha , \bar\theta^{\dot\alpha} \}$. 
\par 
Like to Smolin \cite{smolin} we would like to introduce 
an infinitesimal operator 
$\delta\hat{B}(x^\mu \rightarrow z^\mu = x^\mu + \delta x^\mu)$ 
of a displacement of the 
wormhole mouth 
\begin{eqnarray}
    \delta\hat{B}(x^\mu & \rightarrow & z^\mu) 
    \hat{A}^{\gamma\delta}(y^\mu , x^\mu) = 
    \hat{A}^{\gamma\delta}(y^\mu , z^\mu ) ,
\label{sec5.2-12a}\\
    \hat{A}^{\gamma\delta}(y^\mu , x^\mu) & = & 
    \theta^\gamma \theta^\delta ,
\label{sec5.2-12b}\\
    \hat{A}^{\gamma\delta}(y^\mu , z^\mu) & = & 
    {\theta '}^\gamma {\theta '}^\delta  = 
    \left( \theta^\gamma + \varepsilon^\gamma \right)
    \left( \theta^\delta + \varepsilon^\delta \right) 
    \approx  
    \theta^\gamma \theta^\delta + 
    \varepsilon^\gamma \theta^\delta - 
    \varepsilon^\delta \theta^\gamma 
\label{sec5.2-12c}
\end{eqnarray}
here $\varepsilon^\alpha$ is an infinitesimal Grassmannian number. 
Therefore we have the following equation for the definition of 
$\delta\hat{B}(x^\mu \rightarrow z^\mu)$ operator 
\begin{equation}
    \delta\hat{B}\left(x^\mu \rightarrow z^\mu \right) 
    \theta^\gamma \theta^\delta = 
    \theta^\gamma \theta^\delta + 
    \varepsilon^\gamma \theta^\delta - 
    \varepsilon^\delta \theta^\gamma .
\label{sec5.2-12d}
\end{equation}
This equation has the following solution \cite{dzh01}
\begin{equation}
    \delta\hat{B}\left(x^\mu \rightarrow z^\mu \right) =  
    1 + \varepsilon^\alpha \frac{\partial}{\partial \theta^\alpha} - 
    i\varepsilon^\alpha \sigma^\mu_{\alpha\dot{\beta}} 
    \bar{\theta}^{\dot{\beta}} \partial_\mu - 
    \bar{\varepsilon}^{\dot\alpha} 
    \frac{\partial}{\partial \bar{\theta}^{\dot\alpha}} + 
    i\theta^\alpha \sigma^\mu_{\alpha\dot{\beta}} 
    \bar{\varepsilon}^{\dot{\beta}} \partial_\mu .
\label{sec5.2-12e}
\end{equation}
This allows us to say that 
$\theta = \{ \theta^\alpha , \theta^{\dot\alpha} \}$ 
are the Grassmanian numbers which we should use as some additional 
coordinates for the description of the spacetime foam. 
In this approach the superspace gravity with the anticommuting 
coordinates $\theta$ describes in some approximation 
the spacetime foam. 

\section{Conclusions}

Thus, here we have proposed the approximate model for the
description of the spacetime foam. This model is based on the
assumption that the whole spacetime is 5 dimensional
but $G_{55}$ is the dynamical variable only in the QWHs.
In this case 5D gravity has the solution which we have used as a
model of the individual quantum wormhole. In the approximation 
when the 5D throat of each QWH is contracted 
to a point the spacetime foam can be approximately described by 
a spinor field or Grassmanian anticommuting coordinates on the 
superspace. 
\par
Such model leads to the very interesting experimental consequences.
We see that the spacetime foam has 5D structure and it connected
with the electric field. This observation allows us to presuppose
that the very strong electric field can open a door into 5
dimension! The question is: as is great should be this field ? The
electric field $E_i$ in the CGSE units and $e_i$ in the ``geometrized''
units can be connected by formula
\begin{eqnarray}
e_i & = & \frac{G^{1/2}}{c^2}E_i =
\left (
2.874 \times 10^{-25} \; cm^{-1}/gauss
 \right )E_i ,
\label{cn1}\\
\left [ e_i \right ] & = & cm^{-1} ,
\quad
\left [E_i\right ] = V/cm
\label{cn2}
\end{eqnarray}
As we see the value of $e_i$ is defined by some characteristic
length $l_0$. It is possible that $l_0$ is a length of the
$5^{th}$ dimension. If $l_0 = l_{Pl}$ then
$E_i \approx 10^{57} V/cm$ and this field strength is in the
Planck region, and is will beyond experimental capabilities
to create. But if $l_0$ has a different value it can lead to
much more realistic scenario for the experimental capability
to open door into $5^{th}$ dimension.
\par
Another conclusion of this work is that : 
\textit{the supergravity theories can be considered as approximative
and effective models of the spacetime foam}.

\section{Acknowledgment}

I am grateful for financial support from the Georg Forster Research
Fellowship of the Alexander von Humboldt Foundation and
H.-J. Schmidt for an invitation to Potsdam University.

\end{document}